\documentclass[12pt]{article}
\usepackage{color, amsfonts,amssymb,amsmath,amsthm,graphicx}

\begin{document}

\title{Casimir interaction of two dielectric half spaces
 with Chern-Simons boundary layers}

\author{Valery N. Marachevsky \thanks{email: maraval@mail.ru} \\
{\it St.Petersburg State University,}\\{\it 7/9 Universitetskaya nab.,}\\
{\it St.Petersburg, 199034  Russia} }

\date{}
%
%

\maketitle

\begin{abstract}
A diffraction problem for a flat Chern-Simons layer at plane
boundary of a dielectric half space is solved. The Casimir energy of
two dielectric half spaces  with Chern-Simons layers at
plane-parallel boundaries separated by a vacuum slit is derived.
Crossing from the repulsive to the attractive Casimir force is
analyzed for two Au and two Si half spaces with boundary
Chern-Simons layers. Boundary quantum Hall layers in external
magnetic field lead to Casimir repulsion at nanoscales. We discuss
features that make systems with boundary quantum Hall layers unique
for force measurements and search of long-range interactions beyond
electromagnetism.

\end{abstract}

\section{Introduction}

The Casimir effect \cite{Casimir} is a quantum interaction effect
between macroscopic objects \cite{CasPol} - \cite{Vassilevich}.
Chern-Simons terms \cite{CS} were intensively studied in quantum
field theory \cite{Jackiw1} - \cite{Marino}. The Casimir interaction
of two flat Chern-Simons layers in vacuum was considered in
\cite{Pismak1, Mar17}. The Casimir-Polder potential of a neutral
anisotropic atom in the presence of a flat Chern-Simons layer was
found in \cite{Chern}, charge-parity violating effects due to
Chern-Simons layer interacting with an atom were considered in
\cite{VMStefan}.

The force between two Chern-Simons layers in vacuum is either
attractive or repulsive at all distances depending on values of
constants defining Chern-Simons layers \cite{Pismak1, Mar17}. The
situation changes if there are Chern-Simons layers at plane-parallel
boundaries of dielectric half spaces separated by a vacuum slit of
the width $L$, this problem is considered in the present paper. For
coinciding Chern-Simons layers at plane-parallel boundaries we
demonstrate there can be a minimum of the Casimir energy and, as a
result, Casimir repulsion at nanoscales.

It is natural to perform the Casimir force measurements with quantum
Hall layers as Chern-Simons boundary layers. Suppose one puts the
system in a constant external magnetic field perpendicular to the
layers. The constant $a$ of the Chern-Simons action is uniquely
defined by the plateau of the quantum Hall effect in terms of QED
fine structure coupling $\alpha$ and an integer or a fractional
number $n$ characterizing this plateau as $a = \alpha n$
\cite{QHE1}. The plateau of the quantum Hall effect is defined by
the magnitude of an external magnetic field. Without external
electric field parallel to the quantum Hall layer there are no
spatial components of the Hall current in the boundary layer, it is
reasonable to perform Casimir force measurements in this regime.

The Casimir force between two half spaces with Chern-Simons boundary
layers can be evaluated from the first  principles and with high
precision: constants of Chern-Simons layers are defined by the
plateaus of the quantum Hall effect, frequency dependent dielectric
permittivities of the underlying semispaces can be extracted from
the tabulated optical data \cite{Palik}. The total force in the
Casimir experiments typically consists of two parts: the Casimir and
the electrostatic contributions \cite{Exp1, Exp2}. The difference in
the theoretical and experimental determination of the force may be a
signal of long-range forces beyond electromagnetism measured in
experiment. Long-range interactions beyond electromagnetism arise
due to exchange of light scalar particles predicted in many
extensions of the Standard Model leading to Yukawa-type potentials
\cite{Pot1}, Yukawa-type potentials in the extra-dimensional
theories with a low-energy compactification scale \cite{Pot2, Pot3},
power-type potentials \cite{Pot4, Pot5}, axion exchange potential
\cite{Pot6}. In the interaction range above a few micrometers the
strongest constraints on potentials of long-range interactions
follow from Cavendish type experiments \cite{Pot7, Pot8, Pot9,
Pot10}, at shorter separations the strongest constraints on
Yukawa-type potentials and axion-to-nucleon coupling constants were
obtained from various Casimir effect experiments \cite{Casbook,
Exp1, Exp2, Pot11}.

To sum up, there are several features that make systems with quantum
Hall surface layers unique for force measurements: stabilization of
force measurements at short separations due to Casimir repulsion,
quick decrease of the Casimir force in system with Chern-Simons
layers in comparison to the Lifshitz force for two dielectric half
spaces \cite{Lifshitz} at separations approaching the minimum of the
Casimir energy, precise evaluation of the Casimir force from the
first principles of quantum field theory, possibility of changing
plateaus in the quantum Hall effect regime by external magnetic
field resulting in a change of the force at a given separation.
Geometry of two half spaces with plane-parallel boundaries separated
by a vacuum slit $L$ is the simplest geometry where novel features
due to Chern-Simons boundary layers occur.

We proceed as follows. In Sec.$2$ we solve a diffraction problem for
Chern-Simons layer at the boundary of dielectric half space
characterized by a permittivity $\varepsilon(\omega)$, reflection
and transmission coefficients for diffraction of an electromagnetic
plane wave are derived. In Sec.$3$ we apply scattering formalism
\cite{Marachevskyreview}, \cite{Mar1} - \cite{Mar16} to derive
formulas for the Casimir energy of two dielectric half spaces with
Chern-Simons layers at plane-parallel boundaries separated by a
vacuum slit $L$. We study Casimir forces for Au and Si half spaces
with Chern-Simons boundary layers, crossing from the repulsive to
the attractive Casimir force is analyzed.

We use units $\hbar=c=1$.

\section{Diffraction problem}
The action with Chern-Simons layer at $z=0$ has the form:
\begin{equation}
S=\frac{a}{2} \int \varepsilon^{z\nu\rho\sigma} A_\nu F_{\rho\sigma}
dt dx dy \label{CS1}
\end{equation}
with the current $J^\nu= a \varepsilon^{z \nu \rho \sigma}
F_{\rho\sigma}$ and vector-potential $A_{\nu}$. Equations of
electromagnetic field in the presence of Chern-Simons action
(\ref{CS1}) can be written as follows:
\begin{equation}
\partial_\mu F^{\mu \nu} + a\, \varepsilon^{z \nu \rho\sigma}
F_{\rho\sigma} \delta(z) =0. \label{motion}
\end{equation}

 Consider a flat
Chern-Simons layer put at $z=0$ on a dielectric half space $z<0$
characterized by a frequency dependent dielectric permittivity
$\varepsilon(\omega)$, the magnetic permeability $\mu = 1$. Boundary
conditions on the components of the electromagnetic field can be
written as follows \cite{boundarycondition}:
\begin{align}
E_z|_{z=0^+} - \varepsilon(\omega) E_z|_{z=0^-} &= - 2 a H_z|_{z=0}, \label{cont1}\\
H_x|_{z=0^+} - H_x|_{z=0^-} &=  2 a E_x|_{z=0}, \label{cont2}\\
H_y|_{z=0^+} - H_y|_{z=0^-} &=  2 a E_y|_{z=0}. \label{cont3}
\end{align}

Consider TE ($s$-polarized) electromagnetic plane wave diffracting
from a Chern-Simons layer located at $z=0$ on a dielectric half
space ($z<0$) defined by a dielectric permittivity
$\varepsilon(\omega)$ (the factor $\exp(i\omega t + i k_y y)$ is
dropped for simplicity of notations):
\begin{align}
E_x &= \exp(-ik_z z) + r_{s} \exp(i k_z z) , z>0\\
E_x &= t_{s} \exp(-i k_z^{(2)} z) ,  z<0 \\
H_x &= r_{s \to p} \exp(i k_z z) , z>0  \\
H_x &= t_{s \to p} \exp(- i k_z^{(2)} z) , z<0 .
\end{align}
Here $k_z= \sqrt{\omega^2 - k_y^2}, k_z^{(2)} =
\sqrt{\epsilon(\omega)\omega^2- k_y^2}$.

From the condition (\ref{cont2}) it follows
\begin{equation}
r_{s \to p} - t_{s \to p} = 2 a\, t_{s} . \label{sys1}
\end{equation}
From $E_x|_{z=0^+} = E_x|_{z=0^-}$ we obtain
\begin{equation}
1 + r_{s} = t_{s}.
\end{equation}
From the condition $E_y|_{z=0^+} = E_y|_{z=0^-}$ and Maxwell
equation $E_y = - \frac{1}{i\omega\varepsilon(\omega)} \partial_z
H_x$ it follows that
\begin{equation}
r_{s \to p} k_z = - \frac{k_z^{(2)}}{\varepsilon(\omega)} t_{s \to
p}.
\end{equation}
From the condition (\ref{cont3}) and Maxwell equation $H_y =
\frac{1}{i\omega} \partial_z E_x$ we get
\begin{equation}
k_z(-1 + r_{s}) + k_z^{(2)} t_{s} =  2 a \,
\frac{k_z^{(2)}}{\varepsilon(\omega)} t_{s \to p}. \label{sys4}
\end{equation}
Solving equations (\ref{sys1})-(\ref{sys4}) we find reflection and
transmission coefficients for TE plane wave:
\begin{align}
&r_{s} = \frac{r_s^{f} - a^2 T}{1 + a^2 T}  , &  &t_{s} =
\frac{t_s^f}{1 + a^2 T} ,
\nonumber\\
&r_{s \to p} =  \frac{a T}{1 + a^2 T} , &  &t_{s \to p} = - \frac{a
T}{1 + a^2 T} \frac{\varepsilon(\omega) k_z}{k_z^{(2)}},
\end{align}
where
\begin{equation}
 T = \frac{4 k_z k_z^{(2)}}{(k_z+ k_z^{(2)})
(\varepsilon(\omega) k_z + k_z^{(2)})}
\end{equation}
and
\begin{equation}
r_s^f = \frac{k_z - k_z^{(2)}}{k_z + k_z^{(2)} } , \,\,\,\, t_s^f =
\frac{2k_z}{k_z+k_z^{(2)}}
\end{equation}
are TE Fresnel coefficients for diffraction on a flat dielectric
semispace.

Consider TM ($p$-polarized) electromagnetic plane wave diffracting
from a Chern-Simons layer located at $z=0$ on a dielectric half
space ($z<0$) defined by a frequency dependent dielectric
permittivity $\varepsilon(\omega)$ :
\begin{align}
H_x &= \exp(-ik_z z) + r_{p} \exp(i k_z z) , z>0\\
H_x &= t_{p} \exp(-i k_z^{(2)} z) ,  z<0 \\
E_x &= r_{p \to s} \exp(i k_z z) , z>0  \\
E_x &= t_{p \to s} \exp(- i k_z^{(2)} z) , z<0 .
\end{align}
From the condition $E_x|_{z=0^+} = E_x|_{z=0^-}$ it follows
\begin{equation}
r_{p \to s}=t_{p \to s}. \label{sys5}
\end{equation}
From the condition $E_y|_{z=0^+} = E_y|_{z=0^-}$ and equation $E_y =
-\frac{1}{i\omega\varepsilon(\omega)} \partial_z H_x$ we get
\begin{equation}
k_z(1 - r_{p}) = \frac{k_z^{(2)}}{\varepsilon(\omega)} t_{p}.
\end{equation}
From (\ref{cont2})
\begin{equation}
1+ r_{p} - t_{p} = 2 a \, r_{p \to s}.
\end{equation}
From the condition (\ref{cont3}) and Maxwell equations $H_y =
\frac{1}{i\omega}\partial_z E_x$, $E_y = -
\frac{1}{i\omega\varepsilon(\omega)}\partial_z H_x$ we obtain
\begin{equation}
k_z r_{p \to s} + k_z^{(2)} t_{p \to s} = 2 a \, t_{p}
\frac{k_z^{(2)}}{\varepsilon(\omega)}. \label{sys8}
\end{equation}
Solving equations (\ref{sys5})-(\ref{sys8}) we find reflection and
transmission coefficients for TM plane wave:
\begin{equation}
r_{p} = \frac{r_p^{f} + a^2 T}{1 + a^2 T} , \,\,\,\, t_{p} =
\frac{t_p^f}{1+a^2 T}, \,\,\,\, r_{p \to s} = t_{p \to s} =
\frac{aT}{1+a^2 T} ,
\end{equation}
where
\begin{equation}
r_p^f = \frac{\varepsilon(\omega) k_z -
k_z^{(2)}}{\varepsilon(\omega) k_z + k_z^{(2)}}, \,\,\,\, t_p^f =
\frac{2\varepsilon(\omega) k_z}{\varepsilon(\omega) k_z + k_z^{(2)}}
\end{equation}
are TM Fresnel coefficients for diffraction on a flat dielectric
semispace.

\section{Casimir interaction}

Consider two half spaces ($z \le 0$ and $z \ge L$) characterized in
their volume by a frequency dependent permittivity
$\varepsilon(\omega)$. Chern-Simons terms with constants $a_1$,
$a_2$ are located at plane-parallel boundaries $z=0$ and $z=L$ of
dielectric half spaces. There is a vacuum slit $0<z<L$ between two
half spaces.

The reflection matrix $R_{down}(a_1) = R(a_1)$ from the $z \le 0$
half space is defined by:
\begin{equation}
R (a_1) =
\begin{pmatrix}
r_{s}  & r_{p \to s} \\
r_{s \to p}  & r_p
\end{pmatrix}
= \frac{1}{1+a_1^2 T}
\begin{pmatrix}
r_s^f - a_1^2 T & a_1 T \\
a_1 T & r_p^f + a_1^2 T
\end{pmatrix} .
\end{equation}

The reflection matrix from the $z \ge L$ half space is defined after
euclidean rotation by
\begin{equation}
R_{up}(a_2) = S R(-a_2) S  ,
\end{equation}
where
\begin{equation}
S =
 \begin{pmatrix}
e^{- L\sqrt{\omega^2 + k_x^2 + k_y^2}} & 0 \\
0 &  e^{- L\sqrt{\omega^2+ k_x^2 + k_y^2 }}
\end{pmatrix}
\end{equation}
is a matrix due to a change of the coordinate system $
 x_1=x, y_1= -y, z_1 = - z + L$ (see for example \cite{Mar17, Mar16}).

Consider the system consisting of two dielectric half spaces with
Chern-Simons layers at their plane-parallel boundaries separated by
a vacuum slit of the width $L$ at zero temperature. Temperature
effects in the Casimir effect can be neglected at distances between
half spaces of the order $10$ nm we are interested in. The Casimir
energy is equal
\begin{align}
E (a_1, a_2, L) &= \frac{1}{2} \iiint \frac{d\omega dk_x
dk_y}{(2\pi)^3} \ln\det(I- R_{up}(a_2)R_{down}(a_1)) = \nonumber \\
&\frac{1}{4\pi^2} \int_0^{+\infty} dr r^2 \ln\det (I - e^{-2Lr}
R(-a_2) R(a_1)). \label{ch1}
\end{align}

\begin{figure}
\centering
\includegraphics[width=17cm]{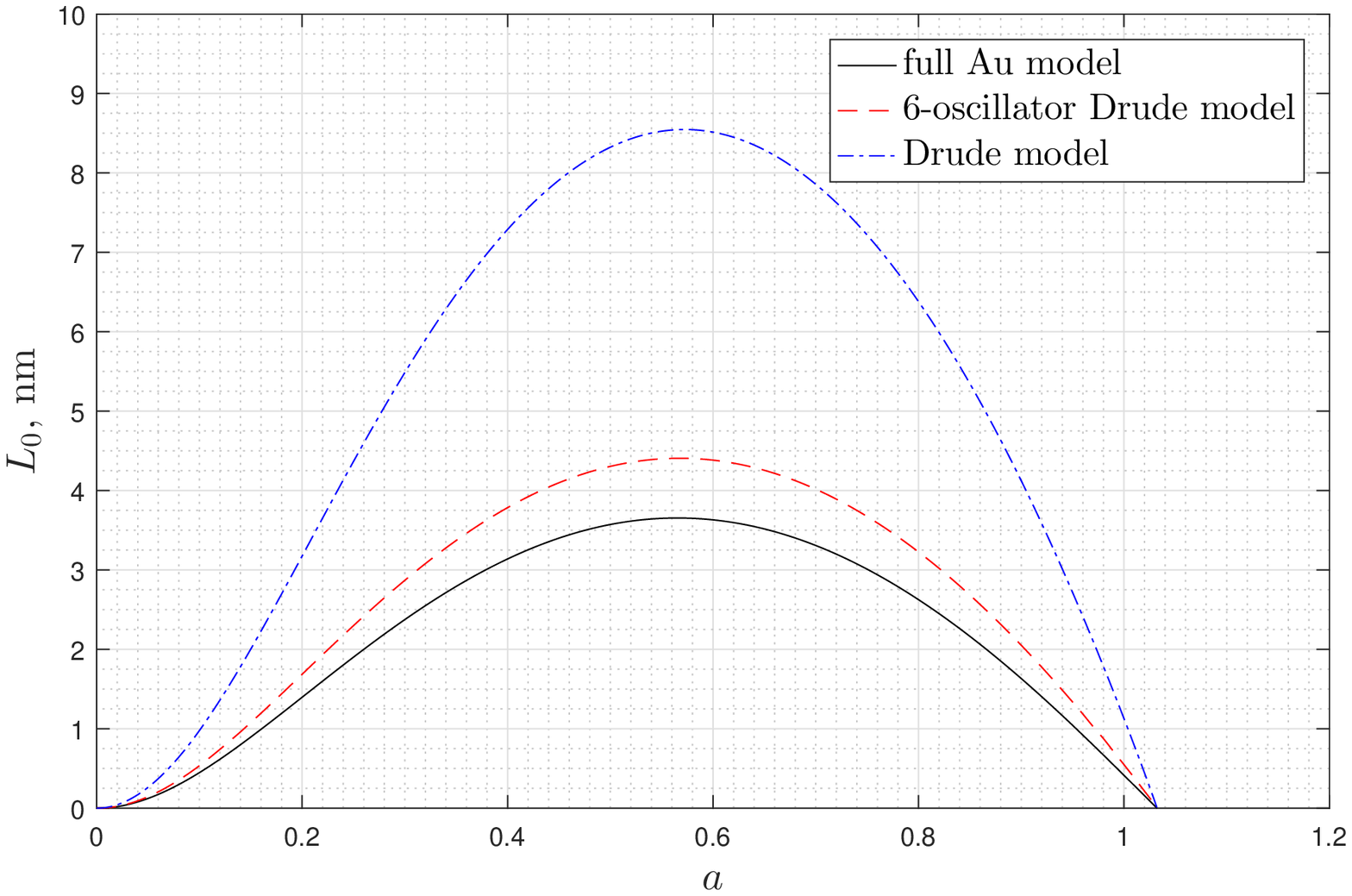}
 \caption{Position of the minimum of the energy $L_0$ for Chern-Simons layers on Au
 semispaces, $a=a_1=a_2$. Results for three models
 of Au dielectric permittivity are shown.}
 \label{figure1}
\end{figure}

\begin{figure}
\centering
\includegraphics[width=17cm]{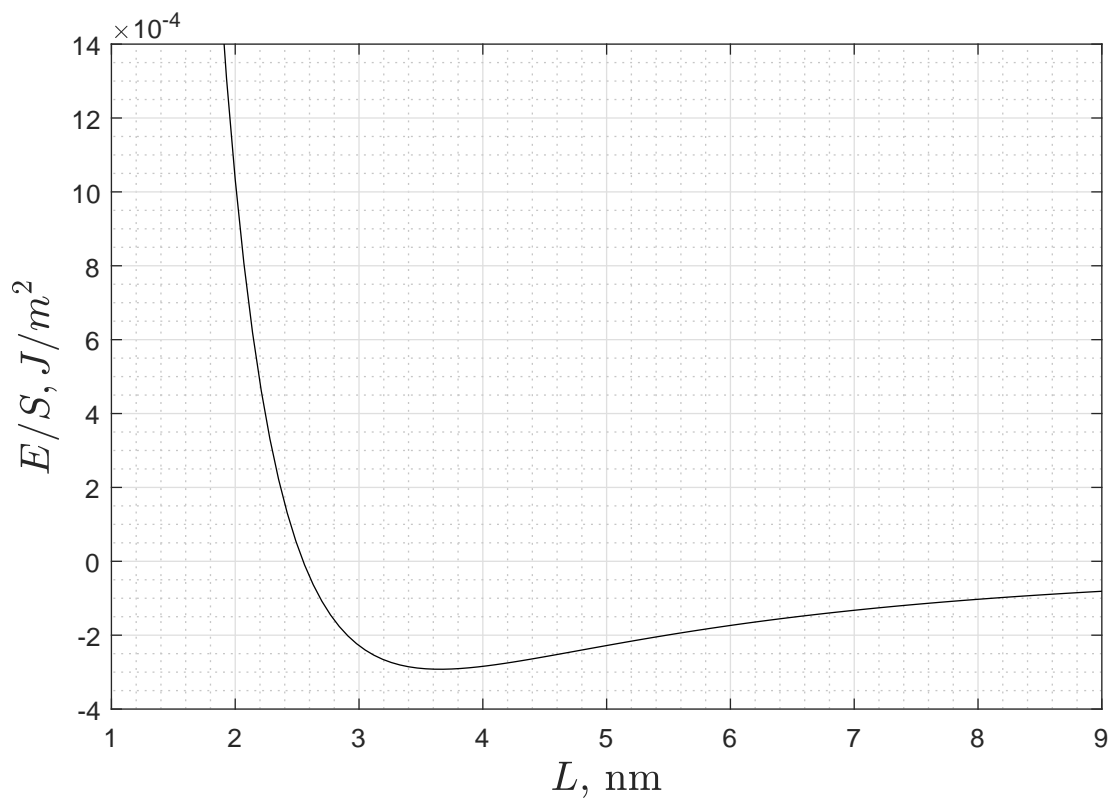}
 \caption{Energy on a unit surface for Chern-Simons layers on Au semispaces obtained
  from full set of known optical data for Au. Chern-Simons constant is $a_1=a_2=0.565$, which corresponds to
 the minimum of energy at $L_0 = 3.65 \,${\rm nm}.  }
 \label{figure2}
\end{figure}

Consider two Au half spaces separated by a vacuum slit with
Chern-Simons boundary layers satisfying the condition $a\equiv
a_1=a_2$. At large separations the Casimir force is attractive. At
short separations there exists a range of parameters $a$ and
distances with a repulsive Casimir force. It is instructive to plot
the position of the energy minimum separating regions of the
repulsive and the attractive force for three different models of
dielectric permittivity: a Drude model $\varepsilon(\omega) = 1 -
\omega_p^2/\omega(\omega+ i\gamma)$ with $\omega_p= 9\, eV, \gamma=
0.035\, eV$, six-oscillator Drude model \cite{Au} and Drude model
with full set of Au data \cite{Palik} taken into account through
Kramers-Kronig relations to evaluate $\varepsilon(i\omega)$. On
Fig.\ref{figure1} dependence of the position of the energy minimum
$L_0$ on the parameter $a$ is shown. The force changes its sign at
separations $L_0 \sim 5$ nm. These distances are typical for the
nonretarded region of the Casimir interaction between two
dielectric/metal half spaces separated by a vacuum slit. To evaluate
the Casimir force in the nonretarded limit one should use optical
data for frequency dispersion of the dielectric permittivity on the
whole frequency axis. This is the reason why the simple Drude model
of Au and even the six-oscillator model of Au can not be used for
precise calculations of the Casimir forces at these separations.
Energy plot corresponding to Chern-Simons parameters $a_1=a_2=0.565$
(corresponding to a maximum value $L_0=3.65$ nm on
Fig.\ref{figure1}) is shown on Fig.\ref{figure2}. The change of the
force from repulsive to attractive behavior takes place at the
distance $L_0=3.65$ nm in this case.

\begin{figure}
\centering
\includegraphics[width=17cm]{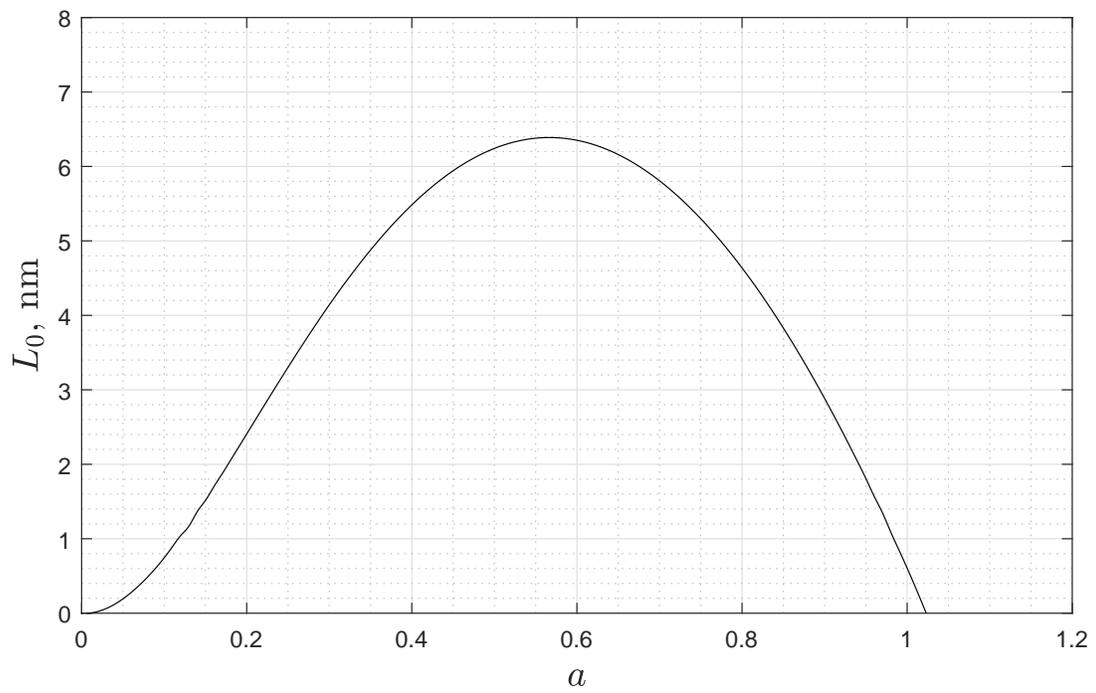}
 \caption{Position of the minimum of the energy $L_0$ for Chern-Simons layers on intrinsic Si semispaces, $a=a_1=a_2$.}
 \label{figure3}
\end{figure}

\begin{figure}
\centering
\includegraphics[width=17cm]{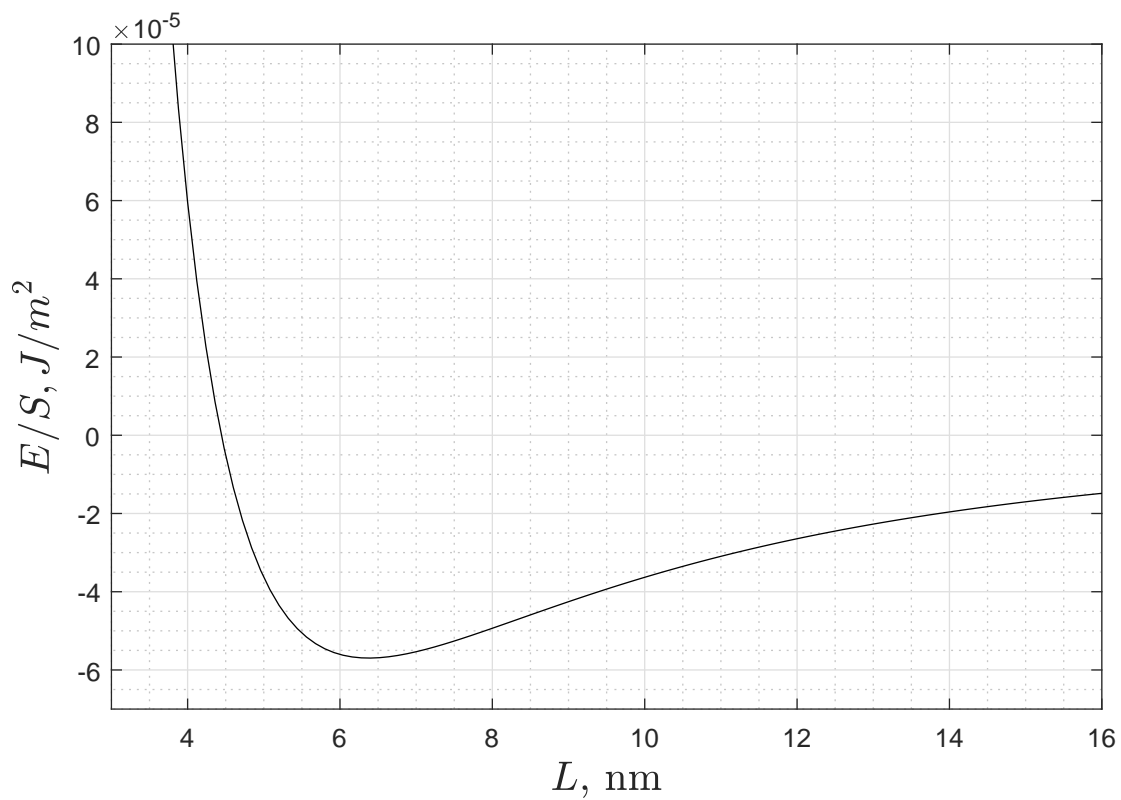}
 \caption{Energy on a unit surface obtained for Chern-Simons layers on intrinsic Si semispaces.
 Chern-Simons constant is $a_1=a_2=0.567$, which corresponds to
 the minimum of the energy at $L_0 = 6.39 \,${\rm nm}.  }
 \label{figure4}
\end{figure}

For dielectric permittivity of intrinsic Si we used the model
\cite{Si}. The change of the force sign for Si corresponding to
$a_1=a_2=0.567$ takes place at the distance $L_0=6.39$ nm
(Fig.\ref{figure3}, Fig.\ref{figure4}), which is about $2$ times
larger than in the case of Au.

\begin{figure}
\centering
\includegraphics[width=15cm]{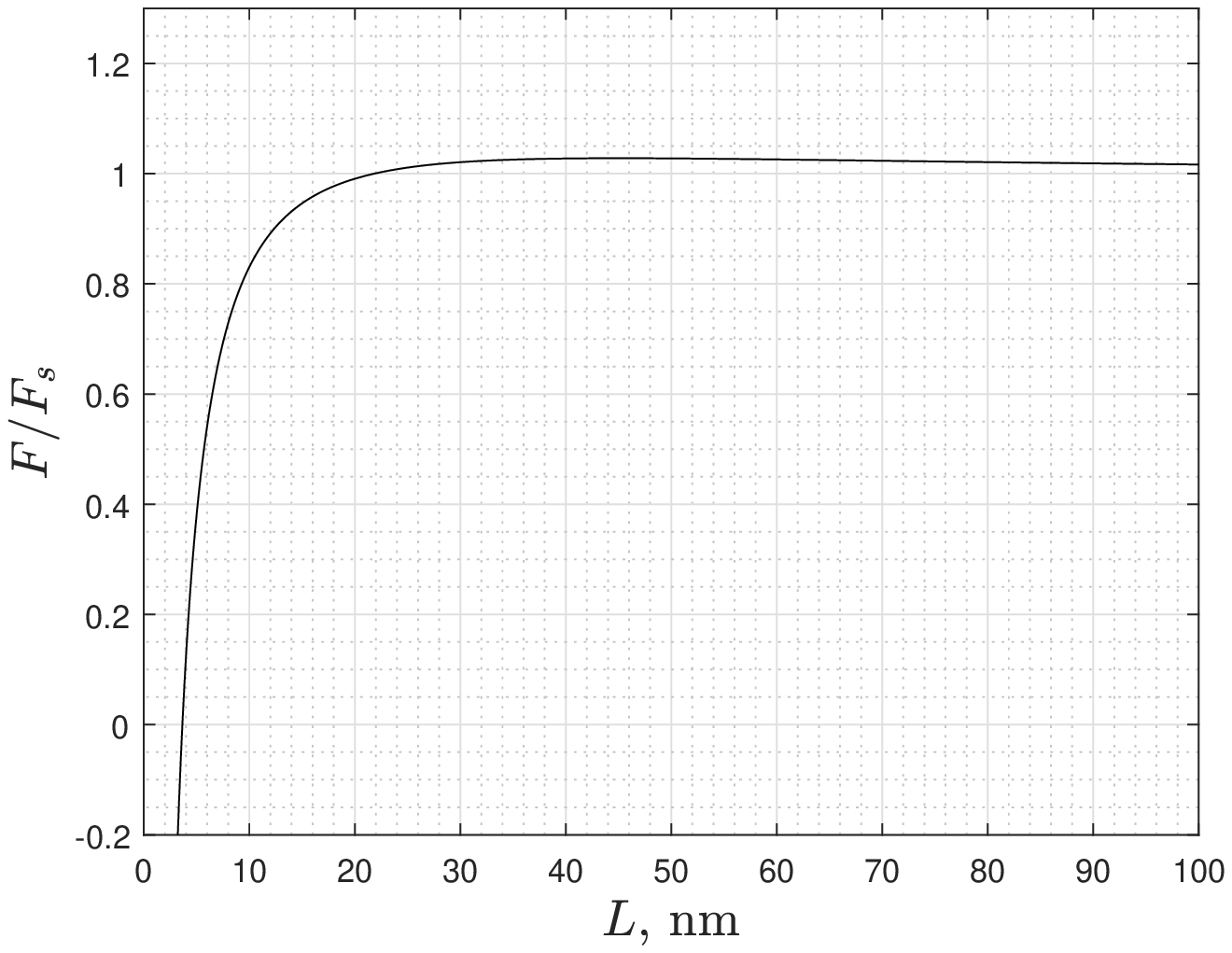}
 \caption{ Ratio of the force $F$ with Chern-Simons layers at the
 boundaries of two Au semispaces to the Lifshitz force $F_s$ between two Au semispaces separated by a distance $L$.
 Chern-Simons constants are $a_1=a_2= 0.565$.}
 \label{figure5}
\end{figure}

\begin{figure}
\centering
\includegraphics[width=15cm]{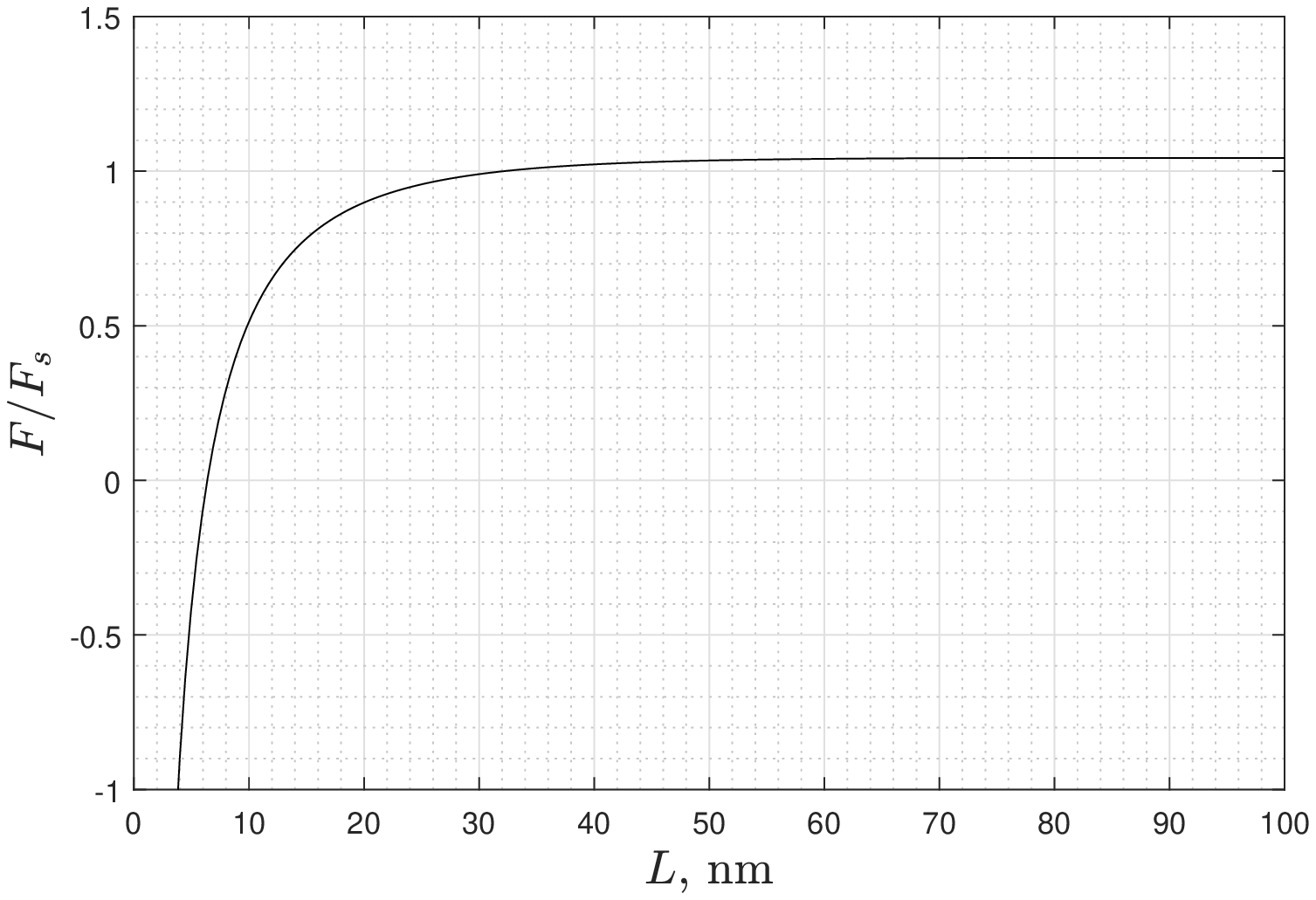}
 \caption{ Ratio of the force $F$ with Chern-Simons layers at the
 boundaries of two intrinsic Si semispaces to the Lifshitz force $F_s$ between two intrinsic Si semispaces separated by a distance $L$.
 Chern-Simons constants are $a_1=a_2= 0.567$.}
 \label{figure6}
\end{figure}

It is also instructive to plot ratio of the Casimir force  with
Chern-Simons layers at the boundaries of two half spaces $F$ to the
Lifshitz force $F_s$ \cite{Lifshitz} (the force between two
dielectric half spaces separated by a distance $L$). These ratios
for Au and Si are shown on Fig.\ref{figure5} and Fig.\ref{figure6}
respectively; transition between repulsive and attractive regimes of
the Casimir force due to Chern-Simons boundary layers is clearly
seen. Note that ratio of the forces quickly decreases at $L_0 < L
\lesssim 20$ nm for Au and $L_0 < L \lesssim 30$ nm for Si systems.

Casimir repulsion at short separations is explained as follows.
Lifshitz force power law effectively changes from attractive
retarded $L^{-4}$ to attractive nonretarded $L^{-3}$ behaviour for
dielectrics (metals) at distances of the order $L \sim 10$ nm. On
the other hand, the force between two Chern-Simons layers in vacuum
has $L^{-4}$ behavior at all separations and thus dominates the
total force at separations of the order $L \lesssim 10$ nm. For an
interval $a \in [0 , a_0]$, where $a_0 \approx 1.032502$, and the
condition $a\equiv a_1=a_2$ the Casimir force between two
Chern-Simons layers in vacuum is always repulsive. As a result, the
sum of the Lifshitz force and the force between two Chern-Simons
layers in vacuum effectively leads to Casimir repulsion at short
separations.

Chern-Simons constants of the layers at $z=0$ and $z=L$ will be
equal ($a\equiv a_1=a_2$) in a constant external magnetic field
perpendicular to the layers if one uses quantum Hall layers at the
boundaries of dielectric half spaces. Quantum Hall layers are
effectively described by Chern-Simons terms at plateaus of the
quantum Hall effect with Chern-Simons constants $a = \alpha n$,
where $\alpha$ is QED fine structure constant, $n$ is an integer or
a fractional number corresponding to integer or fractional quantum
Hall effect. It is reasonable to put external electric field
parallel to the layers equal to zero to avoid spatial components of
Hall currents in the boundary layers. Quantum Hall layers will lead
to Casimir repulsion at nanoscales for $a \lesssim 1$ for Au and Si.
Essential decrease of the Casimir force due to presence of boundary
quantum Hall layers in comparison with Lifshitz force should be seen
at separations $L \lesssim 20$ nm for Au and $L \lesssim 30$ nm for
Si half spaces.

\begin{figure}
\centering
\includegraphics[width=15cm]{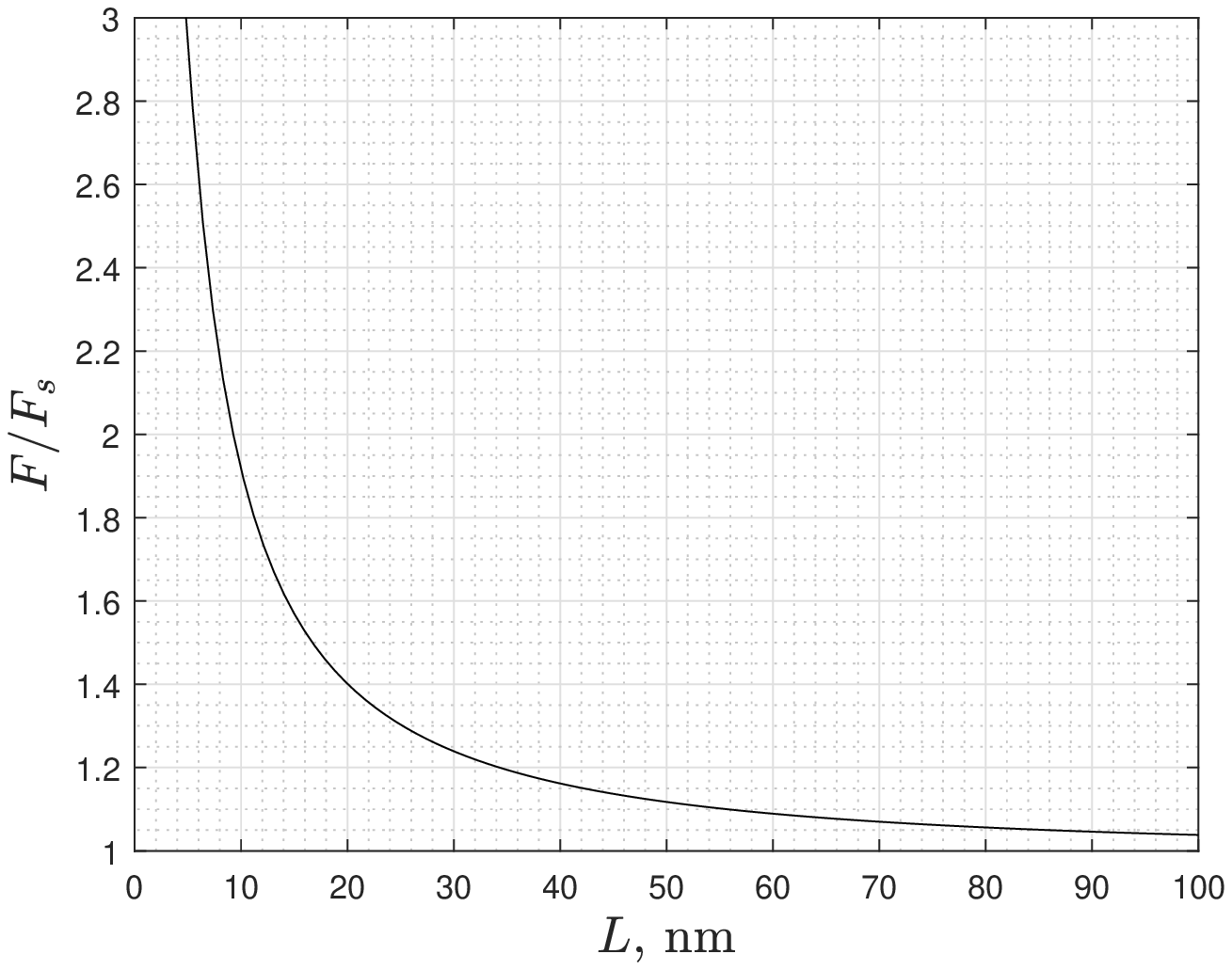}
 \caption{ Ratio of the force $F$ with Chern-Simons layers at the
 boundaries of two Au semispaces to the Lifshitz force $F_s$ between two Au semispaces separated by a distance $L$.
 Chern-Simons constants are $a_1=-a_2= 0.565$.}
 \label{figure7}
\end{figure}

\begin{figure}
\centering
\includegraphics[width=15cm]{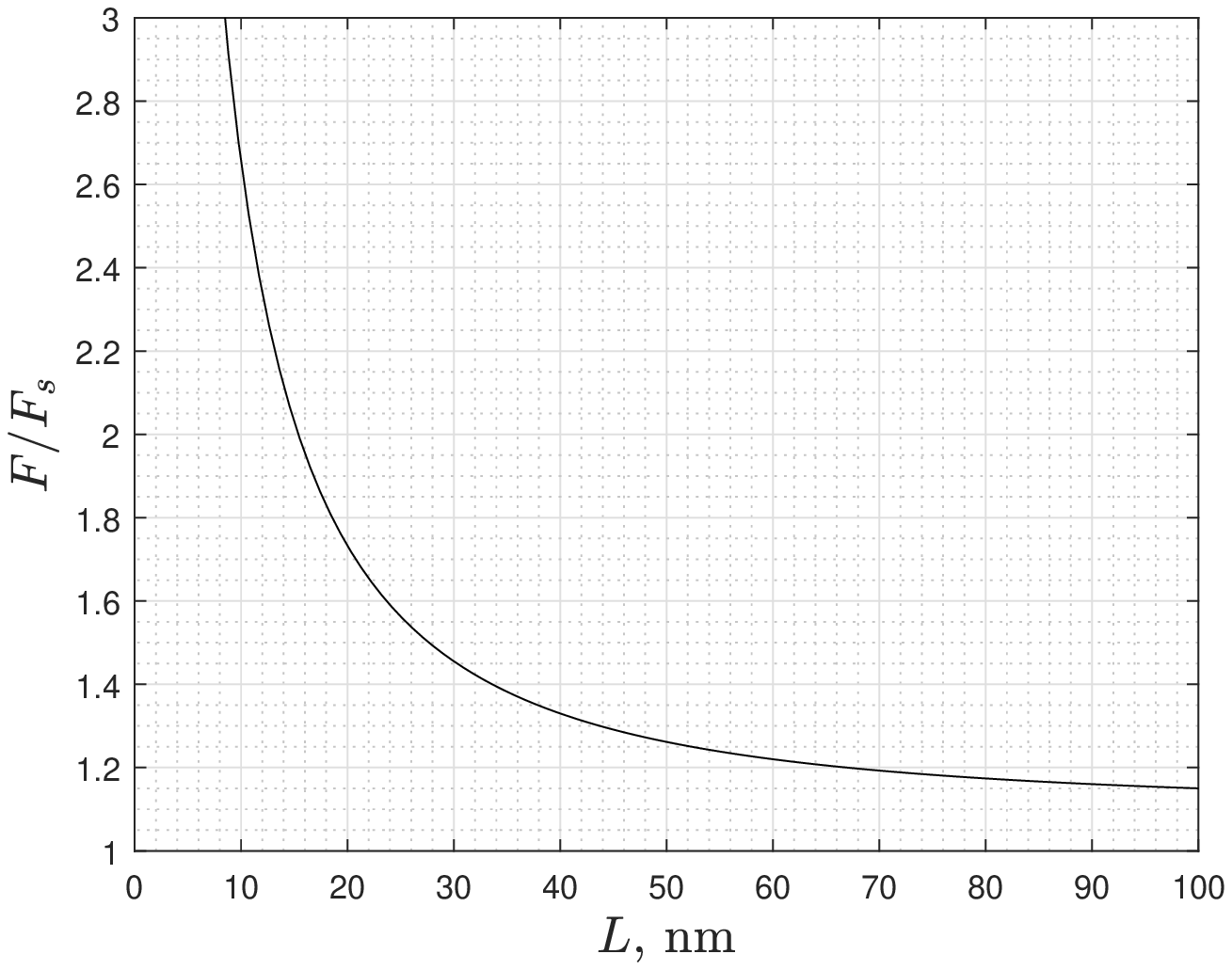}
 \caption{ Ratio of the force $F$ with Chern-Simons layers at the
 boundaries of two intrinsic Si semispaces to the Lifshitz force $F_s$ between two intrinsic Si semispaces separated by a distance $L$.
 Chern-Simons constants are $a_1=-a_2= 0.567$.}
 \label{figure8}
\end{figure}

For $a_1=-a_2$ the force is always attractive, ratios of the Casimir
force  with Chern-Simons layers at the boundaries of two half spaces
$F$ to the Lifshitz force $F_s$ are plotted on Fig.\ref{figure7} and
Fig.\ref{figure8} for Au and Si respectively. The attractive Casimir
force in this case is due to a theorem that opposites obtained by
mirror images of each other attract when they are separated by a
vacuum slit \cite{Klich}.

\section{Conclusions}
In  this paper we present a solution of a diffraction problem for a
Chern-Simons plane layer located at the boundary of a dielectric
half space characterized by a frequency dependent dielectric
permittivity $\varepsilon(\omega)$. Casimir forces for two
dielectric half spaces with Chern-Simons boundary layers separated
by a vacuum slit $L$ have remarkable properties. For equal
Chern-Simons terms the Casimir energy has a minimum at $L=L_0$ for
an interval of constants $a$ of Chern-Simons terms, the Casimir
force for such systems is repulsive at short separations and
attractive at large separations. We have shown that for Au half
spaces with Chern-Simons boundary layers the minimum of the Casimir
energy can be achieved at separation $L_0 = 3.65$ nm, for Si half
spaces with Chern-Simons boundary layers - at $L_0 = 6.39$ nm.
Repulsive behaviour of the Casimir force at short separations should
enhance stability of force measurements at $L \sim L_0$. Quick
decrease of the Casimir force in comparison to the Lifshitz force at
separations approaching the energy minimum is another interesting
feature of systems with Chern-Simons boundary layers that can be
verified in experiments.

It is natural to consider experimental implementation of the
Chern-Simons layers as quantum Hall layers in the presence of
constant external magnetic field perpendicular to the layers and
zero external electric field parallel to the layers. Knowledge of
the Chern-Simons constant $a= \alpha n$ at each plateau of the
quantum Hall effect (characterized by an integer or a fractional
number $n$) is important for precise evaluation of the Casimir force
from the first principles.

Finding the difference between the theoretical force and its
experimental values is a natural way to search for long-range
interactions beyond electromagnetism. Systems with Chern-Simons
boundary layers have unique features  that provide novel
possibilities in experimental search of long-range interactions
beyond electromagnetism via force measurements.

\section{Acknowledgments}
Research was carried out using computational resources provided by
Resource Center “Computer Center of SPbU” (http://cc.spbu.ru/en).
V.M. acknowledges Saint Petersburg State University for a research
grant ${\rm IAS}\_ 11.40.538.2017$.

\newpage

\begin{thebibliography}{99}


\bibitem{Casimir}
H. B. G. Casimir, {\it On the attraction between two perfectly
conducting plates}, {\it Proc. K. Ned. Akad. Wet.} {\bf 51} (1948)
793.

\bibitem{CasPol}
H. B. G. Casimir and D. Polder, {\it The influence of retardation on
the London-van der Waals forces}, {\it Phys.Rev.} {\bf 73} (1948)
360.

\bibitem{Lifshitz}
 E.~M. Lifshitz, {\it The theory of molecular attractive forces between solids},
 {\it Soviet Phys. JETP} {\bf 2} (1956) 73.

\bibitem{Ginzburg1}
Yu. S. Barash and V. L. Ginzburg, {\it Electromagnetic fluctuations
in matter and molecular (Van-der- Waals) forces between them}, {\it
Sov. Phys. Usp.} {\bf 18} (1975) 305.

\bibitem{Ginzburg2}
 Yu. S. Barash and V. L. Ginzburg, {\it Some problems in the
theory of van der Waals forces}, {\it Sov. Phys. Usp.} {\bf 27}
(1984) 467.

\bibitem{Plunien}
G. Plunien, B. M$\rm\ddot{u}$ller and W. Greiner, {\it The Casimir
effect}, {\it Phys.Rept.} {\bf 134} (1986) 87.


\bibitem{Miltonreview}
K. A. Milton, {\it The Casimir effect: recent controversies and
progress}, {\it J.Phys.A: Math. Gen.} {\bf 37} (2004) R 209.

\bibitem{Jaffe}
R. L. Jaffe, {\it Casimir effect and the quantum vacuum}, {\it Phys.
Rev. D} {\bf 72} (2005) 021301(R).

\bibitem{Santangelo}
E. M. Santangelo, {\it Evaluation of Casimir energies through
spectral functions}, {\it Theor. Math. Phys.} {\bf 131} (2002) 527.

\bibitem{BuhmannScheel}
S. Scheel and S. Y. Buhmann, {\it Macroscopic quantum
electrodynamics - concepts and applications}, {\it Acta Phys.
Slovaca} {\bf 58} (2008) 675.

\bibitem{Marachevskyreview}
V. N. Marachevsky, {\it The Casimir effect: medium and geometry},
{\it J.Phys.A: Math. Theor.} {\bf 45} (2012) 374021.

\bibitem{Casbook}
M. Bordag, G. L. Klimchitskaya, U. Mohideen and V. M. Mostepanenko,
{\it Advances in the Casimir Effect}, Oxford University Press,
Oxford (2015).

\bibitem{Buhmann}
S. Y. Buhmann, {\it Dispersion forces, Vols. I, II},
Springer-Verlag, Berlin Heidelberg (2012).

\bibitem{Vassilevich}
D. Fursaev and D. Vassilevich,{\it Operators, Geometry and Quanta:
Methods of Spectral Geometry in Quantum Field Theory},
Springer-Verlag, Dordrecht (2011).

\bibitem{CS}
S. -S. Chern and J. Simons, {\it Characteristic forms and geometric
invariants}, {\it Ann.Math.} {\bf 99}(1) (1974) 48.


\bibitem{Jackiw1}
R. Jackiw,  {\it Fractional charge and zero modes for planar systems
in a magnetic field}, {\it Phys. Rev. D} {\bf 29} (1984) 2375.

\bibitem{Jackiw2}
R. Jackiw and E. J. Weinberg, {\it Self-dual Chern-Simons vortices},
{\it Phys. Rev. Lett.} {\bf 64} (1990) 2234.


\bibitem{Dunne}
G. V. Dunne, {\it Aspects of Chern-Simons theory}, hep-th/9902115 .

\bibitem{Marino}
M. Marino, {\it Chern-Simons Theory and Topological Strings}, {\it
Rev.Mod.Phys.} {\bf 77} (2005) 675.

\bibitem{Pismak1}
V. N. Markov and Yu. M. Pis’mak, {\it Casimir effect for thin films
in QED}, {\it J. Phys. A: Math. Gen.} {\bf 39} (2006) 6525.


\bibitem{Mar17}
V. N. Marachevsky, {\it Casimir effect for Chern-Simons layers in
the vacuum}, {\it Theor.Math.Phys.} {\bf 190}(2) (2017) 315.


\bibitem{Chern}
V. N. Marachevsky and Yu. M. Pis'mak, {\it Casimir-Polder effect for
a plane with Chern-Simons interaction}, {\it Phys.Rev.D} {\bf 81}
(2010) 065005.

\bibitem{VMStefan}
S. Yu. Buhmann, V. N. Marachevsky and S. Scheel, {\it
Charge-parity-violating effects in Casimir-Polder potentials}, {\it
Phys.Rev.A} {\bf 98}(2) (2018) 022510.

\bibitem{QHE1}
Zyun F. Ezawa, {\it Quantum Hall effects field theoretical approach
and related topics}, World Scientific, Singapore (2000).

\bibitem{Palik}
{\it Handbook of Optical Constants of Solids}, edited by E. Palik,
Academic Press, New York (1998).


\bibitem{Exp1}
H. C. Chiu, G. L. Klimchitskaya, V. N. Marachevsky, V. M.
Mostepanenko and U. Mohideen, {\it Lateral Casimir force between
sinusoidally corrugated surfaces: asymmetric profiles, deviations
from the proximity force approximation, and comparison with exact
theory}, {\it Phys.Rev.B \/} {\bf 81} (2010) 115417.

\bibitem{Exp2}
Y.-J. Chen, W. K. Tham, D. E. Krause, D. L$\rm\acute{o}$pez, E.
Fischbach and R. S. Decca, {\it Stronger limits on hypothetical
Yukawa interactions in the 30-8000 nm range}, {\it Phys.Rev.Lett.}
{\bf 116} (2016) 221102.


\bibitem{Pot1}
E. Fischbach and C. L. Talmadge, {\it The Search for Non-Newtonian
Gravity}, Springer, New York (1999).

\bibitem{Pot2}
I. Antoniadis, N. Arkani-Hamed, S. Dimopoulos and G. Dvali, {\it New
Dimensions at a Millimeter to a Fermi and Superstrings at a TeV},
{\it Phys.Lett.B} {\bf 436} (1998) 257.

\bibitem{Pot3}
N.Arkano-Hamed, S.Dimopoulos and G.Dvali, {\it Phenomenology,
astrophysics, and cosmology of theories with submillimeter
dimensions and TeV scale quantum gravity}, {\it Phys.Rev.D} {\bf 59}
(1999) 086004.

\bibitem{Pot4}
L. Randall and R. Sundrum, {\it Large mass hierarchy from a small
extra dimension}, {\it Phys.Rev.Lett.} {\bf 83} (1999) 3370.

\bibitem{Pot5}
L. Randall and R. Sundrum, {\it An alternative to compactification},
{\it Phys.Rev.Lett.} {\bf 83} (1999) 4690.

\bibitem{Pot6}
S. D. Drell and K. Huang, {\it Many-body forces and nuclear
saturation}, {\it Phys.Rev.} {\bf 91} (1953) 1527.

\bibitem{Pot7}
D. J. Kapner, T. S. Cook, E. G. Adelberger, J. H. Gundlach, B. R.
Heckel, C. D. Hoyle and H. E. Swanson, {\it Tests of the
gravitational inverse-square law below the dark-energy length
scale}, {\it Phys. Rev. Lett.} {\bf 98} (2007) 021101.

\bibitem{Pot8}
A. A. Geraci, S. J. Smullin, D. M. Weld, J. Chiaverini and A.
Kapitulnik, {\it Improved constraints on non-Newtonian forces at 10
microns}, {\it Phys.Rev.D} {\bf 78} (2008) 022002.

\bibitem{Pot9}
R. Spero, J. K. Hoskins, R. Newman, J. Pellam and J. Schultz, {\it
Test of the gravitational inverse-square law at laboratory distances
}, {\it Phys.Rev.Lett.} {\bf 44} (1980) 1645.

\bibitem{Pot10}
J. K. Hoskins, R. D. Newman, R. Spero and J.Schultz, {\it
Experimental tests of the gravitational inverse-square law for mass
separations from 2 to 105 cm}, {\it Phys.Rev.D} {\bf 32} (1985)
3084.

\bibitem{Pot11}
V. M. Mostepanenko, {\it A few remarks on the relationship between
elementary particle physics, gravitation and cosmology}, {\it Grav.
Cosmol.} {\bf 22} (2016) 116.


\bibitem{Mar1}
A. Lambrecht and V. N. Marachevsky, {\it Casimir interaction of
dielectric gratings}, {\it Phys.Rev.Lett} {\bf 101} (2008) 160403.

\bibitem{Mar10}
A. Lambrecht and V. N. Marachevsky, {\it Theory of the Casimir
effect in one-dimensional periodic dielectric systems}, {\it
Int.J.Mod.Phys.A} {\bf 24} (2009) 1789.

\bibitem{Jaffe2}
S. J. Rahi, T. Emig, N. Graham, R. L. Jaffe and M. Kardar, {\it
Scattering Theory Approach to Electrodynamic Casimir Forces},
 {\it Phys.\ Rev.\ D} {\bf80} (2009) 085021.

\bibitem{Pirozhenko}
M. Bordag and I. Pirozhenko, {\it Vacuum energy between a sphere and
a plane at finite temperature}, {\it Phys. Rev. D} {\bf 81} (2010)
085023.

\bibitem{bose}
H. Bender, C. Stehle, C. Zimmermann, S. Slama, J. Fiedler, S.
Scheel, S.Y. Buhmann and V.N. Marachevsky, {\it Probing atom-surface
interactions by diffraction of Bose-Einstein condensates}, {\it
Phys.Rev.X} {\bf 4} (2014) 011029.

\bibitem{Mar15}
 I. V. Fialkovsky, V. N. Marachevsky and D. V. Vassilevich, {\it Finite-temperature Casimir effect for graphene},
 {\it Phys.Rev.B} {\bf 84} (2011) 035446.

\bibitem{Mar16}
V. N. Marachevsky, {\it Scattering formalism in the Casimir
interaction of gratings and Chern-Simons layers}, {\it EPJ Web of
Conferences} {\bf 191} (2018) 06014.

\bibitem{boundarycondition}
D. Yu. Pis’mak, Yu. M. Pis’mak and F. J. Wegner, {\it
Electromagnetic waves in a model with Chern-Simons potential}, {\it
Phys.Rev.E} {\bf 92} (2015) 013204.

\bibitem{Au}
G. L. Klimchitskaya, U. Mohideen and V. M. Mostepanenko, {\it
Kramers-Kronig relations for plasma-like permittivities and the
Casimir force}, {\it J.Phys. A: Math.Theor.} {\bf 40} (2007) F339.


\bibitem{Si} A. Lambrecht, I. Pirozhenko, L. Duraffourg and Ph. Andreucci,
{\it The Casimir effect for silicon and gold slabs}, {\it
EuroPhys.Lett.} {\bf 77} (2007) 44006.

\bibitem{Klich}
O. Kenneth and I. Klich, {\it Opposites attract: a theorem about the
Casimir force}, {\it Phys.Rev.Lett.} {\bf 97} (2006) 160401.



\end{thebibliography}
\end{document}